# Direct observation of surface bandgap shrinkage and negative electronic compressibility in $SrTiO_3$

Warakorn Jindata[a,1], Trung-Phuc Vo[b,1], Chutchawan Jaisuk[a,c], Sung-Kwan Mo[d], Thanh-Tien Nguyen[e], Ján Minár[b] and Worawat Meevasana[a,*]

**AFFILIATIONS:**

[a]School of Physics and Center of Excellence on Advanced Functional Materials, Suranaree University of Technology, Nakhon Ratchasima 30000, Thailand

[b]New Technologies-Research Centre, University of West Bohemia in Pilsen, 30100 Pilsen, Czech Republic

[c]EQ Tech Energy Company Limited, Samut Sakhon, 74000, Thailand

[d]Advanced Light Source, Lawrence Berkeley National Laboratory, Berkeley, California 94720, USA

[e]College of Natural Sciences, Can Tho University, Can Tho City 900000, Vietnam

[*]Corresponding author.

*E-mail address:* worawat@g.sut.ac.th

[1]These authors contributed equally to this work.

## ABSTRACT

In this work, we investigate and compare the electronic structures of $SrTiO_3$ and $KTaO_3$ under ultraviolet (UV) light induced electron doping. Using angle-resolved photoemission spectroscopy (ARPES), the evolution of the surface electronic structures of $SrTiO_3$ and $KTaO_3$ is systematically examined as a function of electron density. In contrast to $KTaO_3$, $SrTiO_3$ exhibits a pronounced shrinking of its surface bandgap by approximately 390 meV, accompanied by a counterintuitive shift of the valence band peak toward lower binding energies of up to 200 meV with increasing electron density. This anomalous behavior constitutes a spectroscopic signature of negative electronic compressibility (NEC). Density-functional-theory calculations provide qualitative support for the experimental observations. The calculations show that surface formation already reduces the apparent near-gap separation in $SrTiO_3$, while additional electron accumulation further drives the slab toward a more metallic state; oxygen-vacancy models likewise produce strong bandgap reduction, identifying plausible mechanisms contributing to the observed surface bandgap shrinkage. These findings establish a direct spectroscopic link between bandgap engineering and the NEC effect at the $SrTiO_3$ surface, highlighting the potential of $SrTiO_3$ for next-generation oxide electronic, optoelectronic, and high-performance capacitive energy storage devices applications.

## 1. Introduction

The discovery of a two-dimensional electron gas (2DEG) [1-5] at the interface between the insulating oxides $LaAlO_3$ and $SrTiO_3$ has generated extensive interest in oxide-based low-dimensional electron systems because of their rich and diverse electronic properties, including high carrier mobility [6,7], superconductivity [8,9], large magnetoresistance [10,11], negative electronic compressibility [12], and nonvolatile memory behavior [13]. These findings have established oxide interfaces and surfaces as promising platforms for exploring emergent electronic phenomena and for developing functional oxide electronic devices. Angle-resolved photoemission spectroscopy (ARPES) has played a crucial role in elucidating the electronic structure of 2DEG states in $SrTiO_3$ [14-22]. In particular, we previously demonstrated that a surface 2DEG can be created at the bare $SrTiO_3$ surface without the formation of a heterointerface, and that its carrier density can be systematically increased using ultraviolet (UV) light irradiation [15]. This approach provides an effective and controllable means to tune the surface electronic structure of $SrTiO_3$.

Beyond its fundamental electronic properties, the formation of a surface 2DEG can substantially modify the near-surface bandgap relative to the bulk value due to many-body effects, such as negative exchange interactions in the electron accumulation layer. This behavior opens an opportunity for spectroscopic studies of surface bandgap renormalization [23]. Consequently, controlling and modulating the 2DEG carrier density offers a powerful spectroscopic route to probing bandgap renormalization and many-body interactions in $SrTiO_3$ [24], as well as in other low-dimensional and correlated systems [25-29]. Recently, we employed ARPES to directly observe a bandgap opening in nanoscale highly oriented pyrolytic graphite (nano-HOPG) square patterns, demonstrating the capability of ARPES to investigate electronic structure modifications induced by surface perturbations [30].

The negative thermodynamic density of states, which is described as a counterintuitive decrease in chemical potential (μ) upon increasing the electron density (n); $(1/n^2)(\partial\mu/\partial n) < 0$, can be experimentally observed by using photoemission spectroscopy (PES) techniques [31-35]. From an applied perspective, negative electronic compressibility (NEC) plays a critical role in determining quantum capacitance, thereby offering opportunities for novel oxide-based capacitive and low-power electronic devices [36,37]. Therefore, understanding the relationship between carrier density and chemical potential evolution is essential for both fundamental condensed matter physics and surface-based functional applications.

Here, we use ARPES to systematically investigate and compare the surface electronic structures of $SrTiO_3$ and $KTaO_3$ as a function of UV light-induced electron doping. We report a clear spectroscopic signature of NEC in $SrTiO_3$, manifested by a pronounced surface bandgap shrinkage of approximately 390 meV, accompanied by a counterintuitive shift of the valence band toward lower binding energies of up to 200 meV with increasing electron density. This behavior provides direct evidence of NEC and is not observed in $KTaO_3$ under comparable conditions. These results establish a direct connection between bandgap engineering and the NEC effect at the $SrTiO_3$ surface, highlighting the potential of $SrTiO_3$ surface 2DEGs for applications in electronic, optoelectronic, and high-capacitance energy storage devices.

## 2. Materials and methods

The commercial single-crystal $SrTiO_3$ and $KTaO_3$ samples were used for ARPES measurements. The electronic structure was investigated at Beamline 10.0.1 of the Advanced

Light Source (ALS) using a photon energy of 50 eV for $SrTiO_3$ and 55 eV for $KTaO_3$, respectively. Photoelectrons were analyzed with a Scienta R4000 hemispherical analyzer, providing an energy resolution in the range of 8–35 meV and an angular resolution of 0.35°. The sample was cleaved *in situ* along the (100) crystallographic plane under ultra-high vacuum (UHV) conditions with a base pressure better than $3 \times 10^{-11}$ Torr. ARPES measurements were carried out immediately after cleaving and subsequently under controlled UV light irradiation of the sample surface to modulate the surface 2DEG carrier density.

Bulk $SrTiO_3$ was studied within density functional theory as implemented in Quantum ESPRESSO [38,39]. The calculations employed the cubic five-atom unit cell with a=3.905 Å [40], the PBEsol exchange-correlation functional [41], and the projector augmented-wave pseudopotentials [42] from the PSLibrary [43]. A plane-wave cutoff of 50 Ry and a charge-density cutoff of 240 Ry were used. The Brillouin zone was sampled using 8×8×8 and 12×12×12 Monkhorst-Pack meshes for the self-consistent and non-self-consistent calculations, respectively. Spin-orbit coupling (SOC) was included self-consistently within the noncollinear formalism. The convergence threshold for self-consistency was set to $10^{-4}$ Ry, and a mixing parameter of 0.3 was adopted. The bulk and neutral slab configuration are relaxed with specific conditions and without SOC. Only the topmost two surface layers were relaxed along the z direction, with the in-plane coordinates fixed and the remaining atoms constrained to their bulk-like positions. The selected atoms were allowed to relax along the surface normal according to the calculated forces until all residual force components on the unconstrained atoms were smaller than 0.05 eV/Å.

The $SrTiO_3$ surface was modeled using a 23-atom slab in a supercell with in-plane lattice parameters fixed to a=b=3.905 Å and a cell length of c=43.62 Å along the surface normal, including a vacuum region to suppress interactions between periodically repeated slabs. To model electrostatic electron doping, we adopted a charged-slab approach in which the electron count of the supercell was increased by 0.20 and 0.50 electrons. In the periodic DFT framework, the resulting non-neutral cell is stabilized by a compensating uniform background charge.the resulting excess charge was compensated by the uniform background implemented in Quantum ESPRESSO.

To assess oxygen deficiency as a possible origin of the band-gap shrinkage, we constructed four defective $SrTiO_3$ slab models by removing a single O atom from inequivalent sites in the near-surface region, including the topmost $TiO_2$ layer, the first subsurface SrO layer, and deeper $TiO_2$- and SrO-like layers. In this way, the calculations probe the dependence of the electronic structure on vacancy depth and local bonding environment. The defective slabs were not structurally relaxed after oxygen removal, both to preserve a common reference geometry for all vacancy positions and to keep the calculations within feasible computational cost for the present spin-orbit-coupled slab models. Accordingly, these calculations should be regarded as a controlled qualitative analysis of the primary electronic consequences of oxygen removal, rather than a quantitative treatment of the fully relaxed dilute-defect regime. Given the 1×1 lateral periodicity of the slab, one missing O atom also represents a relatively high vacancy concentration; despite this, the calculations provide a useful test of whether oxygen vacancies are capable of producing the substantial gap reduction observed experimentally.

**3. Results and discussion**

Figure 1 compares the evolution of the surface bandgap in $SrTiO_3$ and $KTaO_3$ as a function of surface electron density ($n_{2D}$) induced by ultraviolet (UV) light irradiation, as revealed by ARPES measurements. Prior to analyzing the bandgap evolution, the dimensionality of the induced electronic states was examined by varying the photon energy and probing the band dispersion along the $k_z$ direction, as shown in Fig. S1. The observed electronic states exhibit negligible $k_z$ dispersion, which is a defining characteristic of two-dimensional electronic systems, confirming the formation of a surface 2DEG. The evolution of the surface bandgap with electron doping was determined from the peak-to-peak and onset-to-onset separation of energy distribution curves (EDCs) between the conduction band (2DEG states) and the valence band dispersions. This approach allows accurate extraction of the conduction band minimum (CBM) and valence band maximum (VBM) positions. As shown in Fig. 1(a), for the onset-to-onset separation, we find that $SrTiO_3$ shows the surface bandgap of approximately 3.27 eV at low electron density, in agreement with the optical bandgap shown in other published report [44]. After increasing the electron density to $7.70 \times 10^{13}$ cm$^{-2}$, the $SrTiO_3$ exhibits a decrease in its surface bandgap to 2.87 eV (see Fig. 1(b)). This behavior corresponds to the peak-to-peak separation, which the $SrTiO_3$ exhibits a pronounced shrinkage of its surface bandgap from 5.41 eV to 5.02 eV as the electron density increases from $1.28 \times 10^{13}$ cm$^{-2}$ to $7.70 \times 10^{13}$ cm$^{-2}$. This behavior indicates to a counterintuitive bandgap reduction with increasing electron density. In contrast, as shown in Figs. 1(c) and 1(d), the $KTaO_3$ displays the opposite trend: its surface bandgap remains unchanged at 3.45 eV and increases from 5.00 eV to 5.16 eV with increasing electron density from $4.07 \times 10^{13}$ cm$^{-2}$ to $7.01 \times 10^{13}$ cm$^{-2}$, as revealed by onset-to-onset and peak-to-peak separations, respectively. Despite their similar perovskite crystal structures, our ARPES measurements provide direct experimental evidence of fundamentally different surface bandgap responses to electron doping in $SrTiO_3$ and $KTaO_3$ under UV light irradiation.

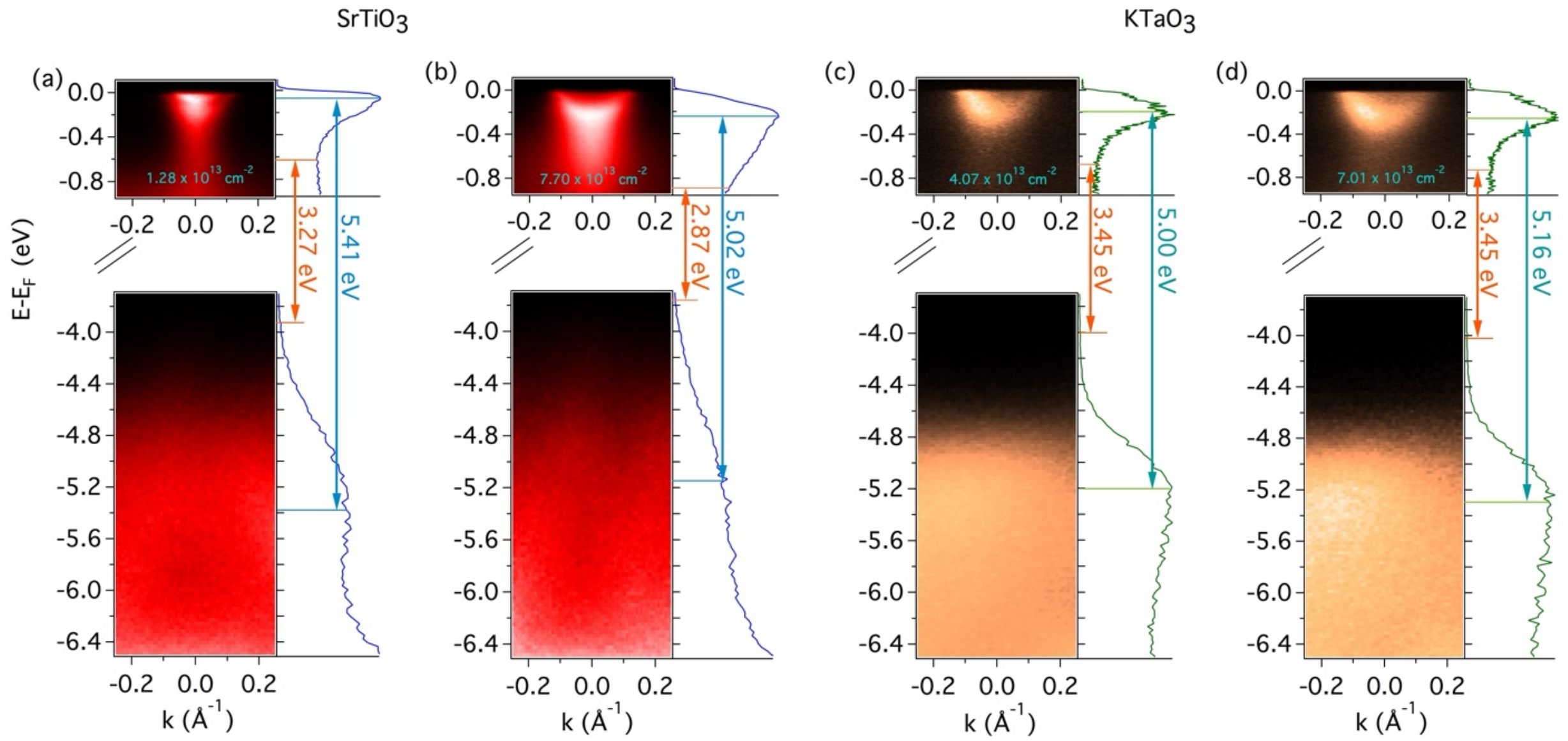

**Fig. 1** ARPES data of the surface electronic structures of $SrTiO_3$ and $KTaO_3$ at different electron densities. (a) and (b), and (c) and (d) show the surface 2DEG and valence band dispersion with corresponding to the energy distribution curves (EDCs) of $SrTiO_3$ and $KTaO_3$ for low and high electron densities, respectively. The EDCs are used to determine the positions of the conduction and valence band onsets (orange arrows), and conduction and valence band peaks (light blue arrows), from which the surface bandgaps are extracted.

To further elucidate the origin of the anomalous bandgap evolution in $SrTiO_3$, we systematically investigated its surface electronic structure as a function of electron doping induced by UV irradiation. Figure 2 presents the evolution of the surface bandgap in $SrTiO_3$ under increasing electron density. The surface electron density $n_{2D}$ was extracted from the Luttinger area using the relation $n_{2D} = (k_F)^2/2\pi$. The maximum surface electron density reaches $7.70 \times 10^{13}$ cm$^{-2}$, consistent with our previous report [15]. Figures 2(a)-2(e) display the ARPES data of $SrTiO_3$ at different surface electron densities, showing the evolution of the surface 2DEG dispersion and the valence band dispersion, together with the corresponding EDCs used to determine the surface bandgap. At low electron density, a single shallow 2DEG band is observed with a Fermi wave vector $k_F = 0.09$ Å$^{-1}$ and a band bottom located approximately -0.06 eV below the Fermi level. With increasing UV irradiation dose, the surface electron density increases gradually, accompanied by a downward shift of the conduction band peak and conduction band onset to higher binding energies, reaching by 190 meV with the ranging from -0.06 eV to -0.25 eV and by 230 meV with the ranging from -0.66 eV to -0.89 eV at the highest electron density (Fig. 2(e)), respectively. Remarkably, the valence band peak and valence band onset exhibit an anomalous shift toward lower binding energies by up to 200 meV with the ranging from -5.47 eV to -5.27 eV and by up to 170 meV with the ranging from -3.93 eV to -3.76 eV as the electron density increases, a signature of the NEC effect [32-35]. The systematic evolution of the conduction band and valence band peak positions as well as the conduction band and valence band onset extracted from the EDCs are summarized as a function of electron density in Fig. 2(f) and Fig. S2(a), respectively. As shown in Figs. 2(g) and S2(b), the peak-to-peak and onset-to-onset separations between the conduction band minimum (CBM) and valence band maximum (VBM) decreases monotonically by approximately 390 meV, from 5.41 eV to 5.02 eV and 400 meV, from 3.27 eV to 2.87 eV, demonstrating a clear shrinkage of the surface bandgap with increasing electron density. The analysis on the change in surface bandgap by onset-to-onset separation presents the same trend with the peak-to-peak separation, which the shrinking value of bandgap is in similar order of magnitude. This opposite shift of the conduction and valence bands cannot be explained by a simple rigid band-filling model. Instead, it provides direct spectroscopic evidence of negative electronic compressibility (NEC), in which the chemical potential decreases with increasing electron density ($\partial\mu/\partial n < 0$) due to strong electron-electron interactions [45,46]. Our ARPES results therefore demonstrate that increasing electron density at the $SrTiO_3$ surface induces a pronounced bandgap shrinkage driven by NEC, highlighting the critical role of many-body effects in shaping the electronic structure of $SrTiO_3$ surface 2DEGs. For comparison, the evolution of the surface bandgap in $KTaO_3$ shows a conventional bandgap increase of approximately 140 meV under peak-to-peak separation (see Figs. S3(a)-S3(d)) and conventional rigid band feature as its surface bandgap remains the same at 3.45 eV under onset-to-onset separation, which are summarized in Figs. S3(f) and S4(b), respectively. In case of bandgap increase, both the conduction band and valence band peaks shift toward higher binding energies by approximately 40 meV and 100 meV, respectively, with increasing electron density (see Fig. S3(e)). In case of rigid band, both the conduction band and valence band onsets shift toward higher binding energies by approximately 90 meV and 90 meV, respectively, with increasing electron density (see Fig. S4(a)). The chemical potential shifts of the valence band onset and valence band peak relative to the Fermi level as a negative for $SrTiO_3$ and a positive for $KTaO_3$ are summarized as a function of electron density in Figs. S5(a)-S5(c) and Fig. S6(a), and Figs. S5(d)-S5(f) and Fig. S6(b), respectively.

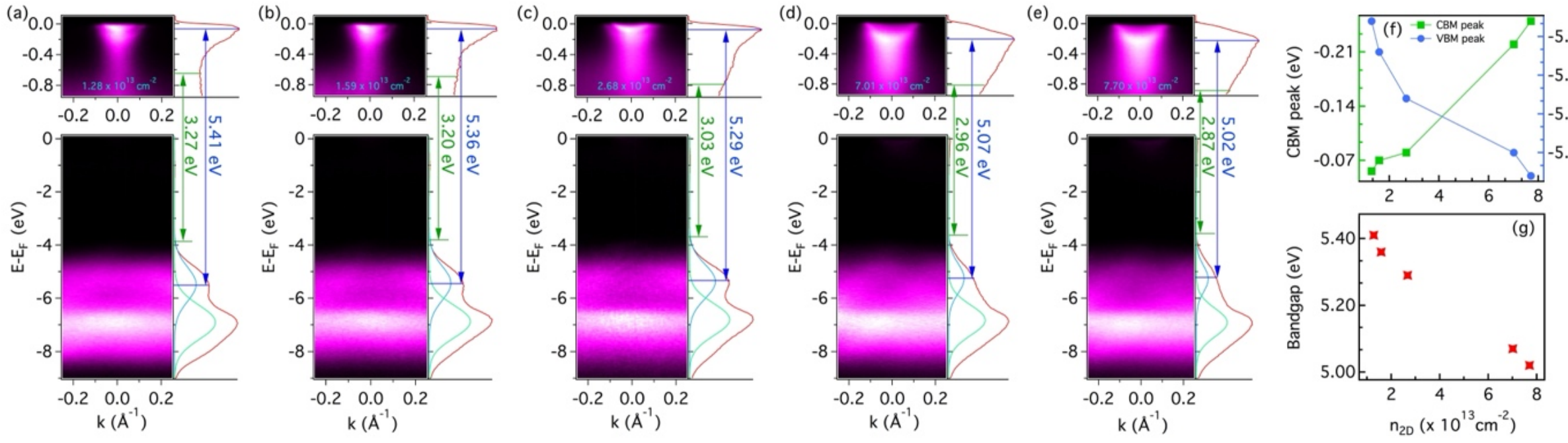

**Fig. 2** ARPES data of the surface electronic structure of $SrTiO_3$ as a function of electron densities. (a)-(e) Surface 2DEG and valence band dispersion presenting with a corresponding EDCs shown on the right, and the extracted conduction and valence band onsets (green arrows), and conduction and valence band peaks (blue arrows) separations are indicated. (f) Summarized the evolution of the conduction band and valence band peaks of $SrTiO_3$ in a function of electron

density. (g) Summarized of the shrinkage feature of surface bandgap of $SrTiO_3$ in a function of electron density.

Figure 3(a) summarizes the surface electron densities extracted from the 2DEG dispersions, revealing a monotonic increase in carrier density with increasing UV irradiation dose and confirming that UV light irradiation provides an effective means to control the surface 2DEG density. The evolution of the chemical potential shift ($\Delta\mu$) as a function of electron density is shown in Fig. 3(b), where the green circles represent values extracted from the valence band peak shifts in Fig. 2. In the presence of NEC, the quantum capacitance $C_q$ is directly related to the thermodynamic density of states through ($Cq = Ae^2\partial n/\partial\mu$) [47], indicating that the quantum capacitance can become negative when $\partial\mu/\partial n < 0$ [37,47]. As shown in Fig. 3(c), the calculated $\partial\mu/\partial n$ exhibits a negative value of approximately -249 × $10^{-12}$ meV $cm^2$ at low electron density. This value is about twice as large as that reported previously for capacitance enhancement in $LaAlO_3/SrTiO_3$ interfaces [47]. Furthermore, $\partial\mu/\partial n$ becomes increasingly negative, reaching approximately 243 × $10^{-10}$ meV $cm^2$ at the maximum electron density of 7.70 × $10^{13}$ $cm^{-2}$, providing compelling evidence for pronounced NEC in the surface 2DEG of $SrTiO_3$. The NEC gives rise to a negative quantum capacitance ($C_q$), which effectively acts in series with the conventional geometric capacitance and can enhance the total capacitance of the system. As reported in $LaAlO_3/SrTiO_3$ interface-based 2DEG, this can drive the capacitance enhancement up to 40% [47]. Therefore, our study is set to pave the way for using a two-dimensional electron system at the bare $SrTiO_3$ surface that is far superior in performance compared to currently used materials in energy and charge storage applications.

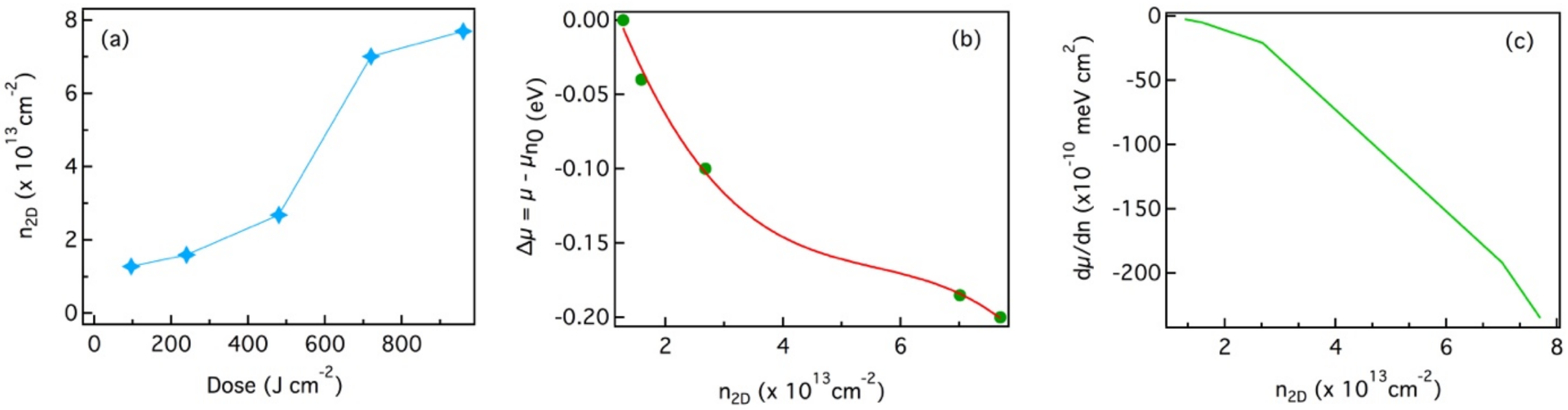

**Fig. 3** (a) The 2DEG charge densities as a function of irradiation dose. (b) The chemical potential shifts ($\Delta\mu$) = $\mu - \mu_{n0}$ as a function of electron density ($n_{2D}$). The green circle symbols are from the measurement data extracted from the valance band peak shift in Fig.2 and the red line is fit. (c) The increase of calculated $\partial\mu/\partial n$ in negative values as a function of $n_{2D}$.

To obtain insight in the mechanism underlying the observed bandgap shrinkage in $SrTiO_3$, we performed the electronic structure calculations shown in Fig. 4. The UV-induced doping may be viewed, at an effective level, as an increase in the electron density localized near the surface [14,15]. Before analyzing explicitly electron-doped configurations, it is therefore important to clarify how the electronic structure is modified simply by going from the bulk crystal to a neutral surface slab. The bulk and neutral-slab band structures thus serve as a natural reference for disentangling intrinsic surface-induced effects from the subsequent changes associated with near-surface electron accumulation. Figure 4 compares the band structures of the bulk and slab models. Owing to the much larger unit cell of the slab, the number of bands increases substantially, resulting in a dense spectrum of states. While the main valence- and conduction-band energy regions remain broadly comparable to those of the bulk system, the reduced symmetry and finite-thickness nature of the slab lead to additional band splitting and a more complex near-gap structure. The calculated bulk $SrTiO_3$ band structure yields an indirect gap of about 1.85 eV, which is consistent with the usual underestimation of the gap within semilocal DFT [48,49], where values around 1.8-1.9 eV are commonly obtained, compared with the experimental indirect and direct gaps of about 3.25 and 3.75 eV [50,51], respectively. In the slab, the apparent gap is further reduced to about 1.0 eV, showing that surface formation alone already produces a substantial modification of the near-gap electronic structure. This bulk-to-slab comparison therefore establishes that a significant part of the gap narrowing can originate from the surface geometry itself, even before additional effects such as excess charge accumulation or oxygen deficiency are considered. Such a reduction is physically reasonable and is close to previous DFT results [52,53].

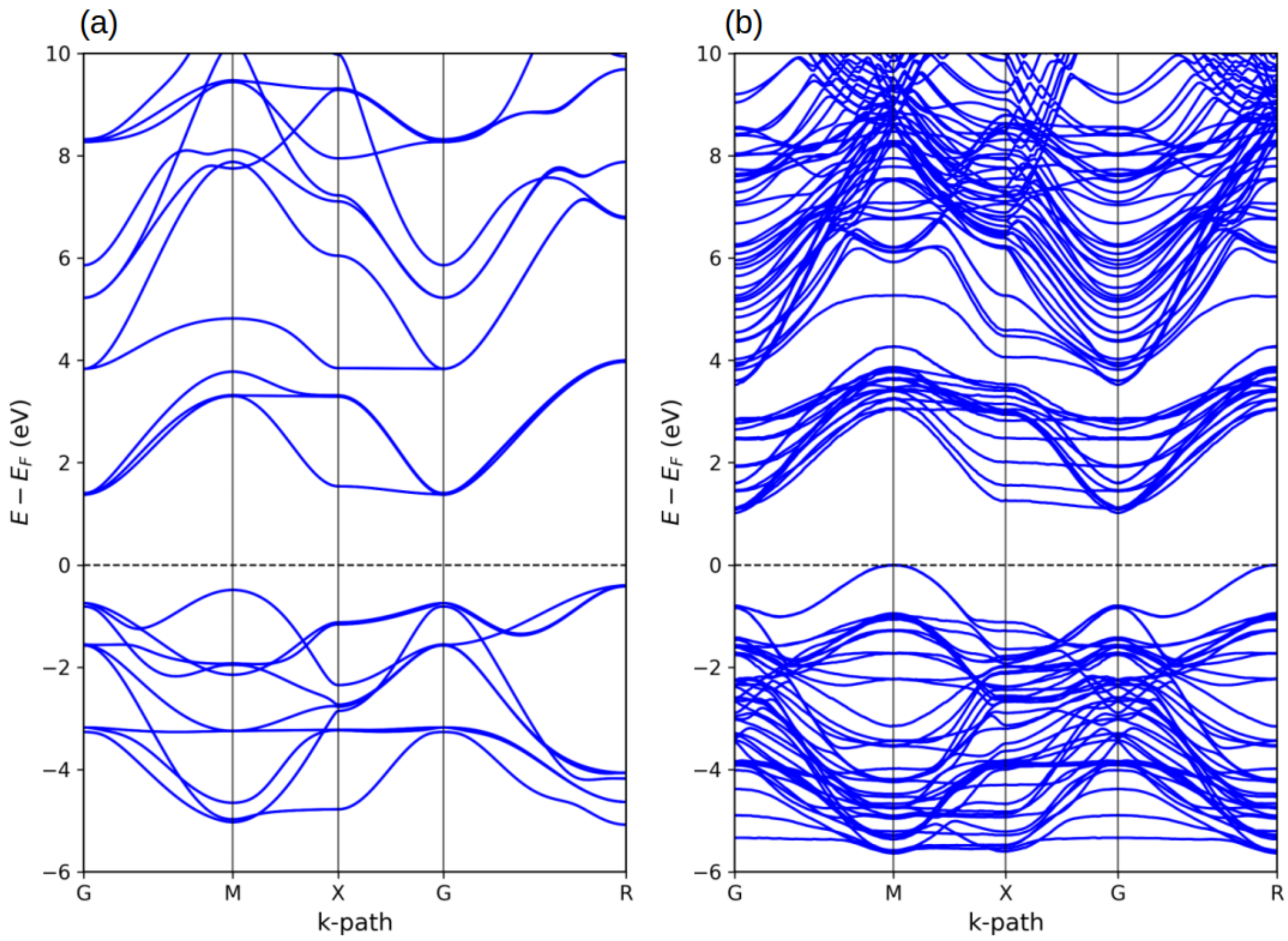

**Fig. 4** Calculated electronic band structures of bulk (a) and slab (b) $SrTiO_3$. The Fermi level is set to zero energy.

Once the intrinsic changes associated with surface formation have been identified from the bulk-to-slab comparison, the next step is to assess the additional role of electron accumulation. To this end, we analyze the band structures of slabs with increasing excess charge, which provide an effective description of progressive near-surface doping and its impact on the low-energy electronic states. The two excess-charge values, q = -0.20 and q = -0.50, were chosen to examine the trend from moderate to stronger electron accumulation at the surface. The corresponding band structures calculated for electron-doped slabs with increasing excess charge are displayed in Fig. 5. In this case, occupations smearing is used to account for the partial occupation of conduction-band-derived states induced by electron doping, whereas for the neutral system (Fig. 4(b)), occupations fixed is employed because the slab remains insulating. The resulting band structures are plotted relative to the Fermi level of each individual calculation. Therefore, the energy zero is not identical among the neutral and doped cases: in the neutral slab, the Fermi level lies within the gap, while in the doped slabs it shifts upward owing to the additional electrons. Consequently, the comparison mainly reflects the evolution of the electronic structure with respect to the chemical potential of each system, rather than an absolute band alignment or band-gap renormalization [54]. Within this convention, increasing electron doping leads to a progressively more metallic character, with conduction-related states approaching and crossing the Fermi level. These metallic states were experimentally confirmed [55]. The trend in Fig. 5 can be described in terms of direct and indirect near-gap separations. The direct separation at Γ, marked by the black arrow, and the indirect separation between the valence-band maximum and conduction-band minimum at different $\vec{k}$ points, marked by the green arrow, both decrease as the excess charge is increased from q = -0.20 to q = -0.50. Because each calculation is plotted relative to its own Fermi level, these quantities should be understood as apparent near-gap separations referenced to the chemical potential. Nevertheless, their systematic reduction clearly indicates that increasing electron accumulation drives the slab toward a more metallic electronic structure. This behavior is consistent with a shrinking effective gap and the onset of metallic character upon electron doping. In this respect, the present results are qualitatively consistent with earlier effective models of surface electron accumulation, although the present charged-slab treatment is more generic than the virtual crystal approximation used in Ref. [56].

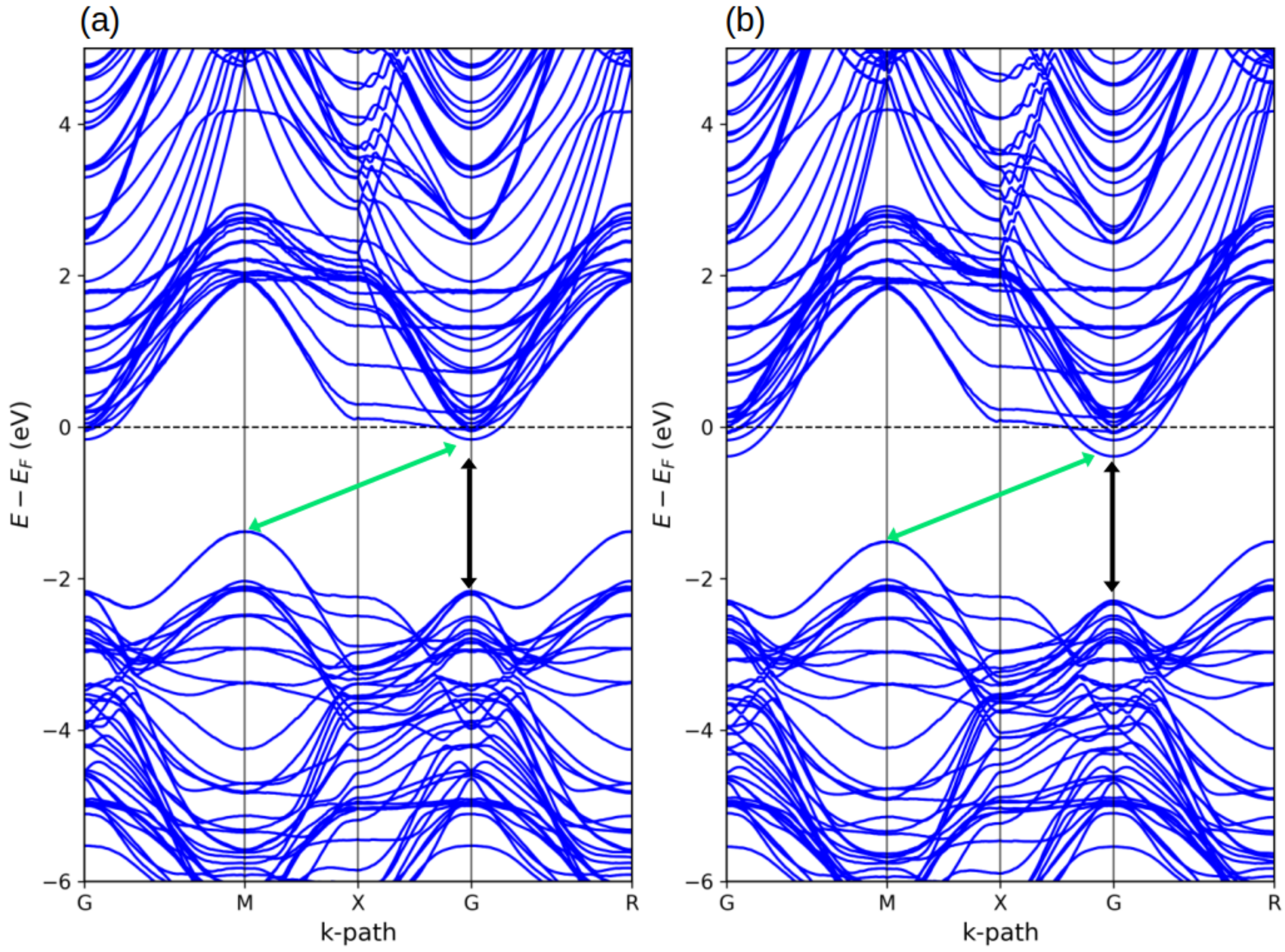

**Fig. 5** Calculated band structures of the doped $SrTiO_3$ slab with additional charges of 0.20 (a) and 0.50 (b) electrons.

In addition to electrostatic electron doping, oxygen vacancies provide another plausible mechanism for the reduction of the band gap in $SrTiO_3$ [57-60]. The band structures in Figure 6 further indicate that this effect depends on the depth of the vacancy. Removing an oxygen atom locally disrupts the Ti-O bonding network, modifies the crystal-field environment of neighboring Ti atoms, and introduces donor-like excess electrons. As a consequence, Ti-derived states can be shifted toward lower energies and defect-related states may emerge near the conduction-band edge or inside the gap, leading to a reduced apparent separation between valence- and conduction-band-derived states (Figs. 6(b) and Fig. 6(d)). In the present 1×1 five-layer supercell, the removal of one oxygen atom corresponds to a relatively high vacancy concentration and therefore represents a strong perturbation of the local electronic structure rather than a dilute-defect limit. As a consequence, the calculated effect is more pronounced than the experimentally observed gap shrinkage: depending on the vacancy position, the band gap is either strongly reduced or completely closed, yielding metallic behavior (Figs. 6(a) and Fig. 6(c)). Although these calculations are not intended to reproduce the experimental gap reduction quantitatively, they clearly demonstrate that oxygen vacancies provide a plausible mechanism for driving gap narrowing and metallization through strong modification of the local Ti–O bonding environment and defect-induced charge redistribution.

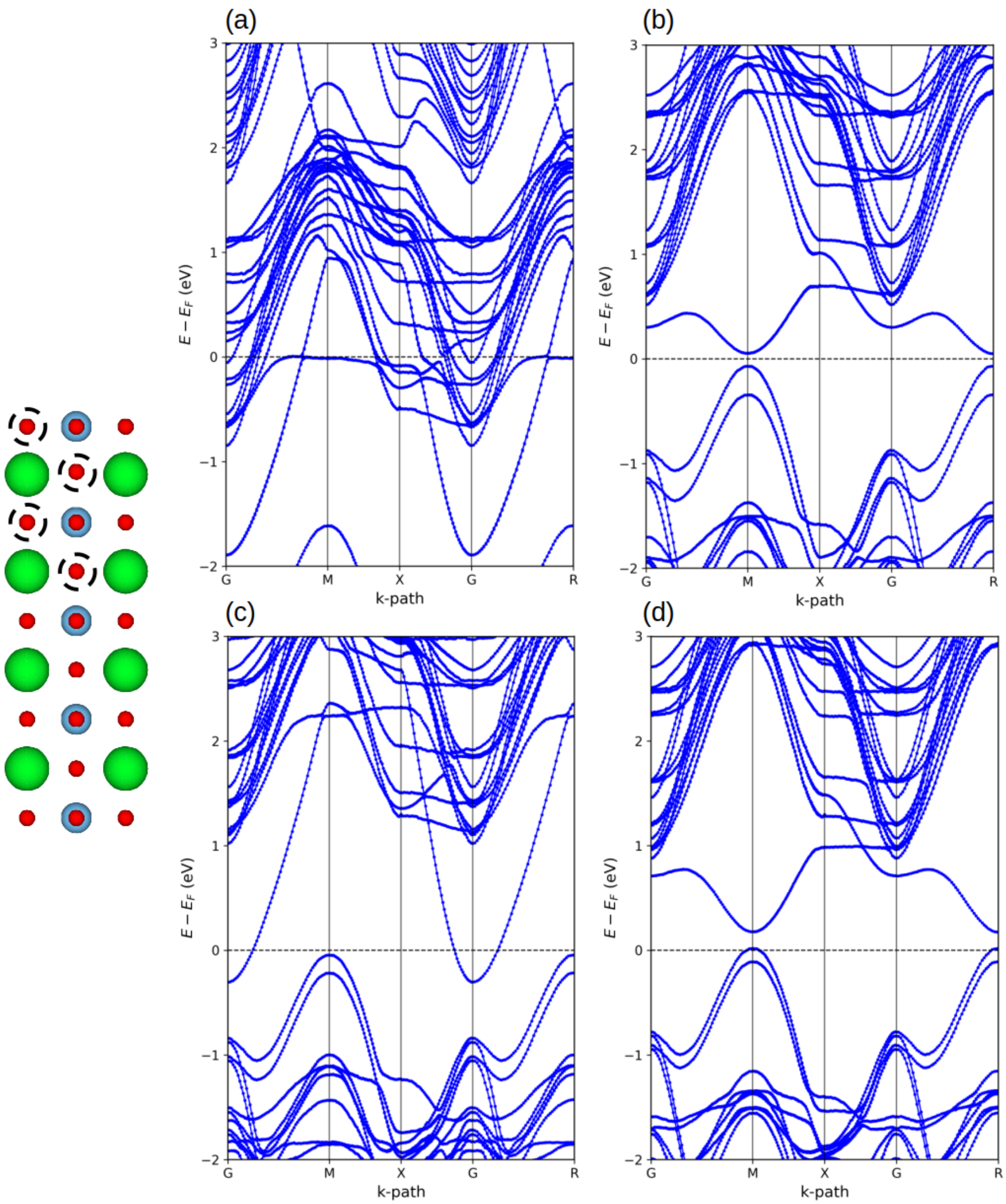

**Fig. 6** Calculated band structures of defective $SrTiO_3$ slabs containing a single oxygen vacancy at four inequivalent sites. The vacancy positions are indicated by dashed circles in the inset. (a)-(d) correspond to oxygen removal from the outermost $TiO_2$ layer, the first subsurface SrO layer, a deeper $TiO_2$ layer, and a deeper SrO-like layer, respectively. In the inset, red, green, and blue spheres represent O, Sr, and Ti atoms, respectively.

We present new evidence revealing distinct differences in the evolution of the electronic structure, specifically the surface bandgap and chemical potential shift, between $SrTiO_3$ and $KTaO_3$. Within a rigid band picture, electron doping is expected to shift the chemical potential toward higher binding energies, while the bandgap remains unchanged or increases, as observed for $KTaO_3$. In contrast, $SrTiO_3$ exhibits anomalous behavior characterized by a pronounced bandgap shrinkage and a counterintuitive shift of the valence band peak toward lower binding energies upon electron doping. This behavior can be understood in terms of negative electronic compressibility (NEC), in which the chemical potential decreases with increasing electron density. Our findings establish a direct spectroscopic link between bandgap engineering and the NEC effect in shaping the surface electronic structure of $SrTiO_3$, highlighting its potential for electronic, optoelectronic, and high-capacitance energy storage device applications.

## 4. Conclusions

In summary, we provide direct experimental evidence of distinct surface electronic structure responses to electron doping in $SrTiO_3$ and $KTaO_3$ using ARPES measurements. $SrTiO_3$ exhibits a clear spectroscopic signature of negative electronic compressibility (NEC), manifested by a counterintuitive chemical potential shift of the valence band peak toward lower binding energies by up to 200 meV with increasing electron density, accompanied by a pronounced surface bandgap shrinkage of approximately 390 meV. In contrast, such behavior is not observed in $KTaO_3$ under comparable doping conditions. Furthermore, the observation of a negative $\partial\mu/\partial n$ demonstrates the emergence of quantum capacitance presence in $SrTiO_3$. Theoretical calculations provide qualitative support for the experimental picture. In particular, the bulk-to-slab comparison shows that surface formation itself already narrows the apparent near-gap separation, charged-slab calculations show a further reduction under increasing electron accumulation, and oxygen-vacancy models produce pronounced bandgap narrowing or even metallic behavior. While the present theoretical treatment is intentionally simplified, it provides qualitative support for the experimental trend and offers a foundation for future studies of additional mechanisms, including surface electric-field effects [61] that may also contribute to the observed bandgap shrinkage. These findings establish a clear connection between bandgap engineering and the NEC effect in the surface 2DEG of $SrTiO_3$, highlighting its potential for optoelectronic applications and, in particular, high-capacitance energy storage devices, where quantum capacitance effects can enable charge storage beyond conventional geometric limits.

**Acknowledgement**

This research has received funding support from the NSRF via the Program Management Unit for Human Resources & Institutional Development, Research and Innovation (Grant No. B13F670064), and EQ Tech Energy Co., LTD. This work was also supported by the project Quantum materials for applications in sustainable technologies (QM4ST), funded as Project No. CZ.02.01.01/00/22_008/0004572 by Programme Johannes Amos Commenius, call Excellent Research (T.-P.V., J.M.) and the Czech Science Foundation Grant No. GA ČR 23-04746S (T.-P.V.).

Supplementary Material for

# Direct observation of surface bandgap shrinkage and negative electronic compressibility in $SrTiO_3$

Warakorn Jindata[a,1], Trung-Phuc Vo[b,1], Chutchawan Jaisuk[a,c], Sung-Kwan Mo[d], Thanh-Tien Nguyen[e], Ján Minár[b] and Worawat Meevasana[a,*]

**AFFILIATIONS:**

[a]School of Physics and Center of Excellence on Advanced Functional Materials, Suranaree University of Technology, Nakhon Ratchasima 30000, Thailand

[b]New Technologies-Research Centre, University of West Bohemia in Pilsen, 30100 Pilsen, Czech Republic

[c]EQ Tech Energy Company Limited, Samut Sakhon, 74000, Thailand

[d]Advanced Light Source, Lawrence Berkeley National Laboratory, Berkeley, California 94720, USA

[e]College of Natural Sciences, Can Tho University, Can Tho City 900000, Vietnam

[*]Corresponding author.

*E-mail address:* worawat@g.sut.ac.th

[1]These authors contributed equally to this work.

## 1. 2D character of the observed band

Figure S1 shows the photon-energy dependence of the ARPES spectra, demonstrating negligible variation of the Fermi momentum ($k_F$). Figures S1(a)-S1(e) present the ARPES energy band dispersion measured along the same momentum cut using low photon flux at photon energies ($E_p$) of 45, 50, 55, 60, and 65 eV, respectively. In all cases, a clear electron-like parabolic dispersion is observed, and the extracted band dispersions (blue solid lines) yield nearly identical $k_F$ positions, as indicated by the arrows. Figure S1(f) summarizes the $k_F$ values obtained from Figs. S1(a)-S1(e), showing that the $k_F$ remains constant within the experimental photon-energy range. This result confirms that the observed electronic states are robust against changes in photon energy and supports their two-dimensional nature.

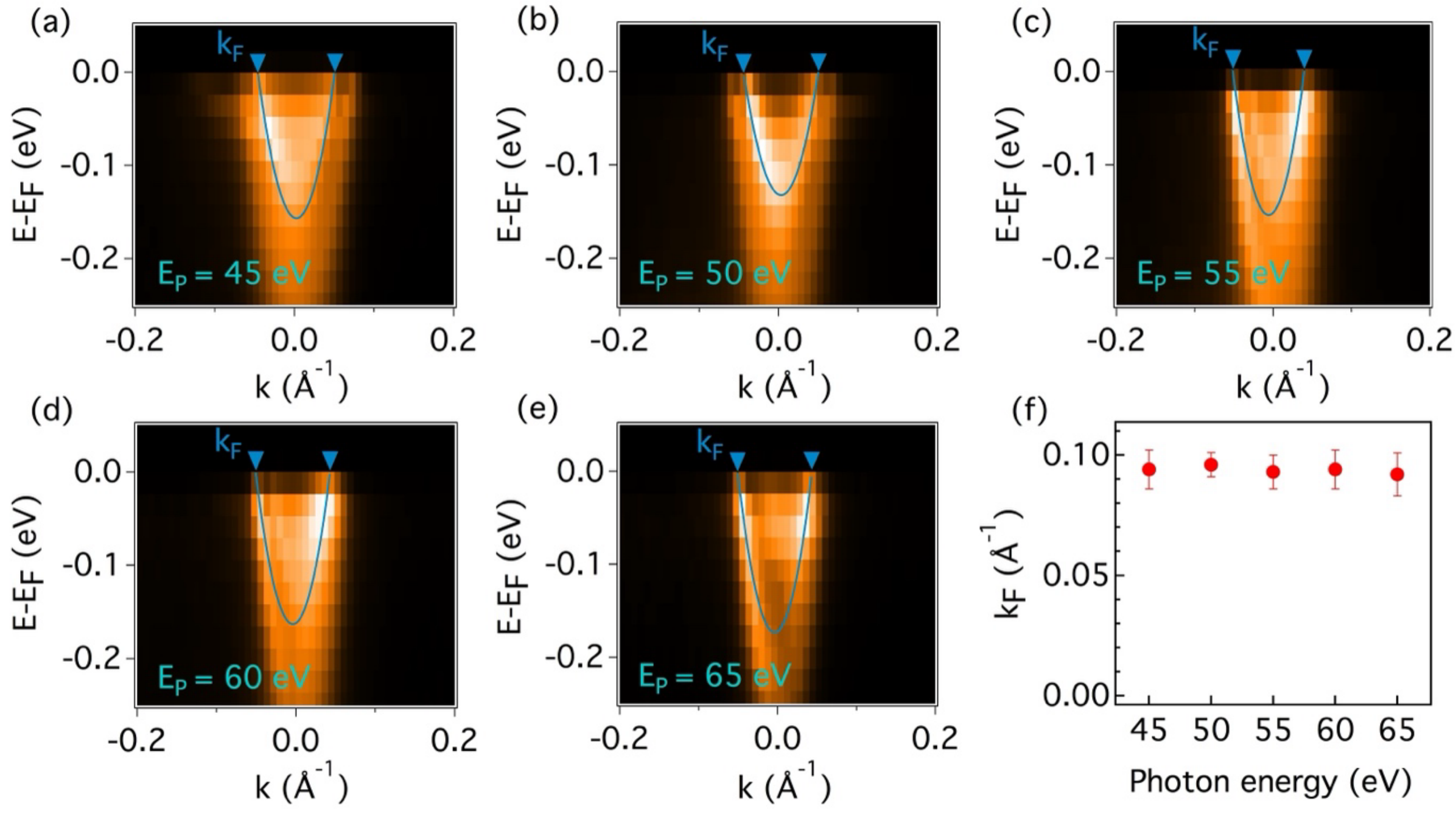

**Fig. S1** Photon energy dependence of ARPES data indicating negligible changes of the Fermi momentum ($k_F$) dispersion. (a) – (e) show ARPES data measured with low photon flux at various photon energies ($E_p$) indicated in the figure. (f) Summarized the $k_F$ extracted from (a) – (e).

**2. Determining the change of bandgap of $SrTiO_3$ upon electron doping by using onset-to-onset separation**

As shown in Figs. S2(a)-S2(b), for the onset-to-onset separation, the evolution of the CBM onset shows a downward shift to higher binding energies, reaching approximately 230 meV with the ranging from -0.66 eV to -0.89 eV at the highest electron density. Intriguingly, the VBM exhibit an anomalous shift toward lower binding energies by up to 170 meV, ranging from -3.93 eV to -3.76 eV with increasing of electron density. The systematic evolution of the CBM onset and VBM onset are summarized as a function of electron density in Fig. S2(a). As shown in Fig. S2(b), the separation between the CBM and the VBM decreases monotonically by approximately 400 meV, from 3.27 eV to 2.87 eV, demonstrating a clear shrinkage of the surface bandgap with increasing electron density (see Figs. 2(a)-2(e)), demonstrating a clear shrinkage of the surface bandgap with increasing electron density. The analysis on the change in bandgap by onset-to-onset presents the same trend with the peak-to-peak method, which the shrinking value of bandgap is in similar order of magnitude with the peak-to-peak separation (390 meV). A signature of negative electronic compressibility (NEC) is still observed in $SrTiO_3$ under analysis the CBM and VBM by onset-to-onset, as the $SrTiO_3$ shows a shrinkage of the surface bandgap by approximately 400 meV, accompanied by a counterintuitive shift of the onset of valence band toward lower binding energies of up to 170 meV with increasing electron density.

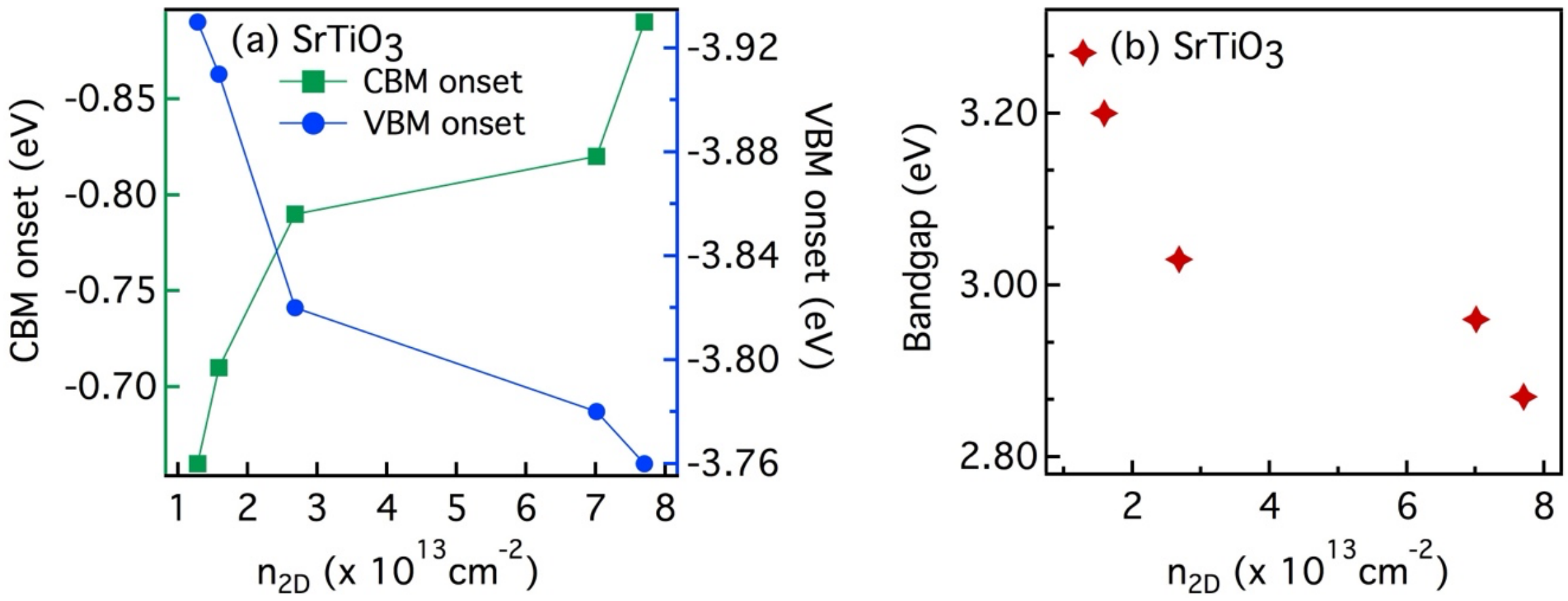

**Fig. S2** (a) Summarized the evolution of the CBM onset and the VBM onset in a function of electron density. (b) Summarized of the shrinkage feature of surface bandgap of $SrTiO_3$ in a function of electron density.

**3. ARPES data of the $KTaO_3$**

Figure S3 presents the evolution of the two-dimensional (surface) bandgap of $KTaO_3$ as a function of electron density. As shown in Figs. S3(a)-S3(d), increasing electron density causes the conduction band (2DEG states) shift progressively toward higher binding energies, indicating enhanced band filling and stronger downward band bending near the surface. Simultaneously, the valence band also shifts toward higher binding energies. By analyzing the peak-to-peak separation between the conduction band and valence band features, the surface bandgap of $KTaO_3$ is found to increase monotonically with electron density, reflecting conventional band-filling behavior. As summarized in Fig. S3(e), quantitative extraction of the band-peak positions shows that the conduction and valence bands shift toward higher binding energies by approximately 40 meV and 100 meV, respectively, as the electron density increases. These shifts result in a total bandgap widening of up to 140 meV, as summarized in Fig. S3(f).

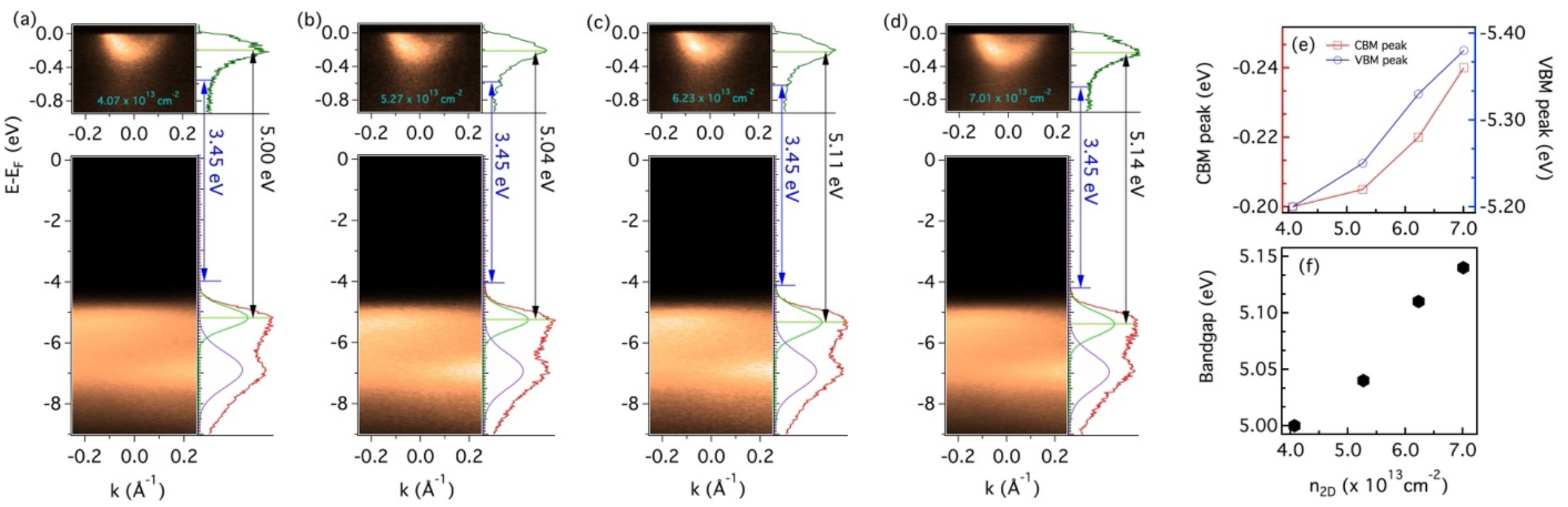

**Fig. S3** ARPES data of the surface electronic structure of $KTaO_3$ as a function of electron densities. (a)-(d) Surface 2DEG and valence band dispersion presenting with a corresponding EDCs shown on the right, and the extracted conduction and valence band onsets (blue arrows) and, conduction and valence band peaks (black arrows) separations are indicated. (e) Summarized the evolution of the conduction band and valence band peaks of $KTaO_3$ in a function of electron density. (f) Summarized of the increase of surface bandgap of $KTaO_3$ in a function of electron density.

As shown in Figs. S4(a)-S4(b), for the onset-to-onset separation, the evolution of the CBM onset shows a downward shift to higher binding energies, reaching approximately 90 meV with the ranging from -0.55 eV to -0.64 eV at the highest electron density. However, the VBM shows a shift toward higher binding energies by up to 90 meV, ranging from -4 eV to -4.09 eV with increasing of electron density. The systematic evolution of the CBM onset and VBM onset are summarized as a function of electron density in Fig. S4(a). As shown in Fig. S4(b), the separation between the CBM onset and VBM onset remains the same at 3.45 eV (see Figs. S3(a)-S3(d)), indicating a rigid band feature, in which the surface bandgap remains unchanged under electron doping.

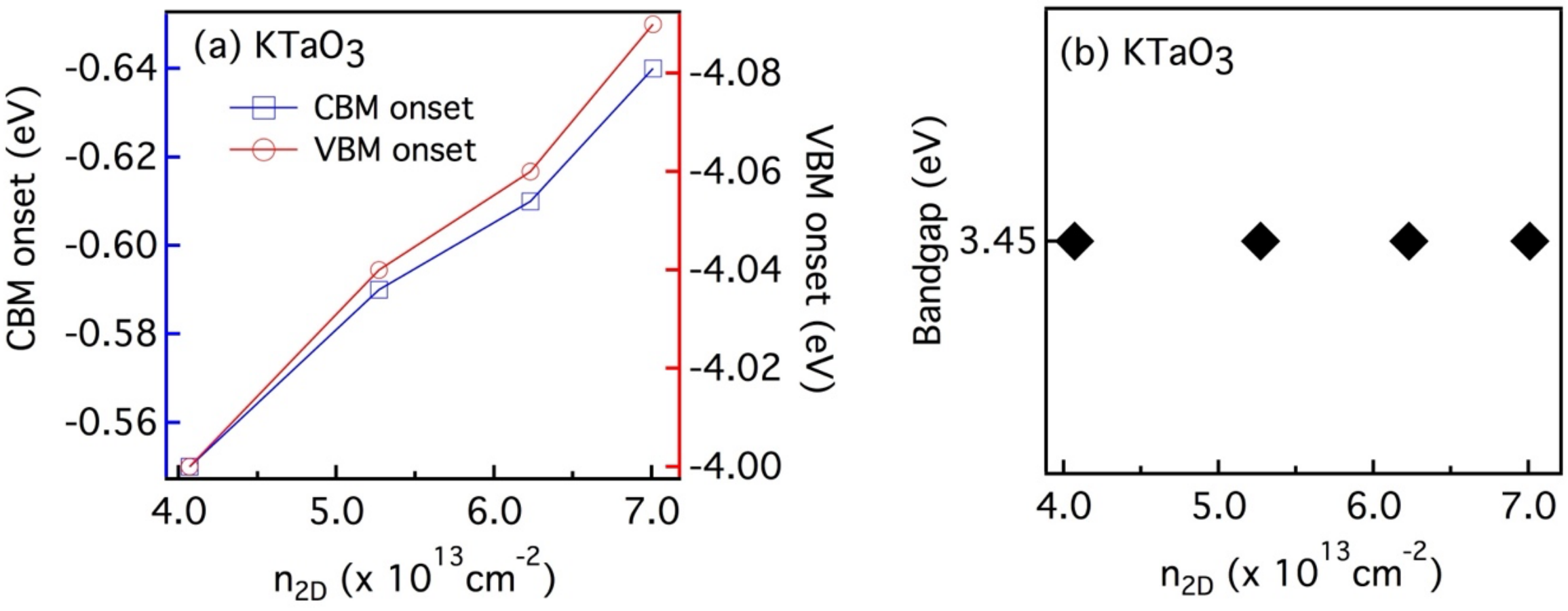

**Fig. S4** (a) Summarized the evolution of the CBM onset and the VBM onset in a function of electron density. (b) Summarized of the rigid band feature of surface bandgap of $KTaO_3$ in a function of electron density.

**4. Determining the valence band maximum (VBM) of $SrTiO_3$ and $KTaO_3$ by using the onset of valence band ARPES spectrum**

Figure S5 compares the evolution of the onset of valence band ARPES spectrum of $SrTiO_3$ and $KTaO_3$ as a function of electron density. As shown in Figs. S5(a) and S5(d), the linear fit is used to determine the VBM of valence band ARPES spectra [1] for $SrTiO_3$ and $KTaO_3$, respectively. For $SrTiO_3$, a counterintuitive shift of the onset of valence band toward lower binding energies is observed with increasing electron density, reaching up to 170 meV (see Fig. S5(b) and zoom-in spectra in S5(c)). In contrast, the $KTaO_3$ exhibits a conventional shift of the onset of valence band toward higher binding energies with increasing electron density by up to 130 meV (see Fig. S5(e) and zoom-in spectra in S5(f)).

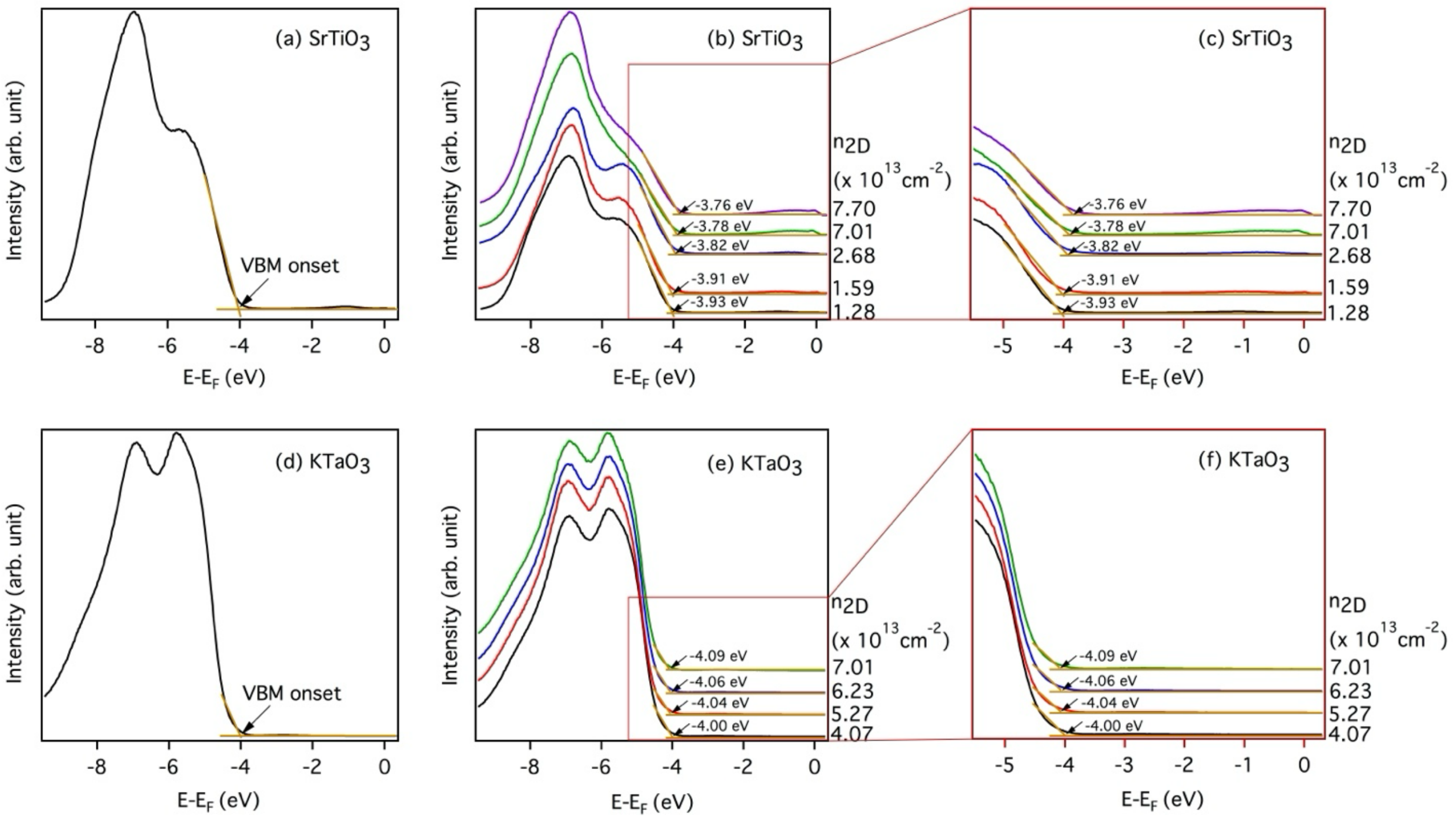

**Fig. S5** (a), (d) The example of linear fitting (gold line) at the onset of valence band ARPES spectrum for determining the valence band maximum (VBM) of $SrTiO_3$ and $KTaO_3$, respectively. (b), (c) and (e), (f) Comparison of the evolution of VBM extracted from the onset of valence band ARPES spectra of (a) $SrTiO_3$ and (b) $KTaO_3$ as a function of electron density.

**5. Determining the valence band maximum (VBM) of $SrTiO_3$ and $KTaO_3$ by using the peak of valence band ARPES spectrum**

Figure S6 compares the evolution of the valence band spectra of $SrTiO_3$ and $KTaO_3$ as a function of electron density. For $SrTiO_3$, as shown in Fig. S6(a), a counterintuitive shift of the valence band toward lower binding energies is observed with increasing electron density, reaching up to 200 meV (see the zoom-in spectra). In contrast, as shown in Fig. S6(b), $KTaO_3$ exhibits a conventional shift of the valence band toward higher binding energies with increasing electron density, with a maximum shift of approximately 40 meV (see the zoom-in spectra).

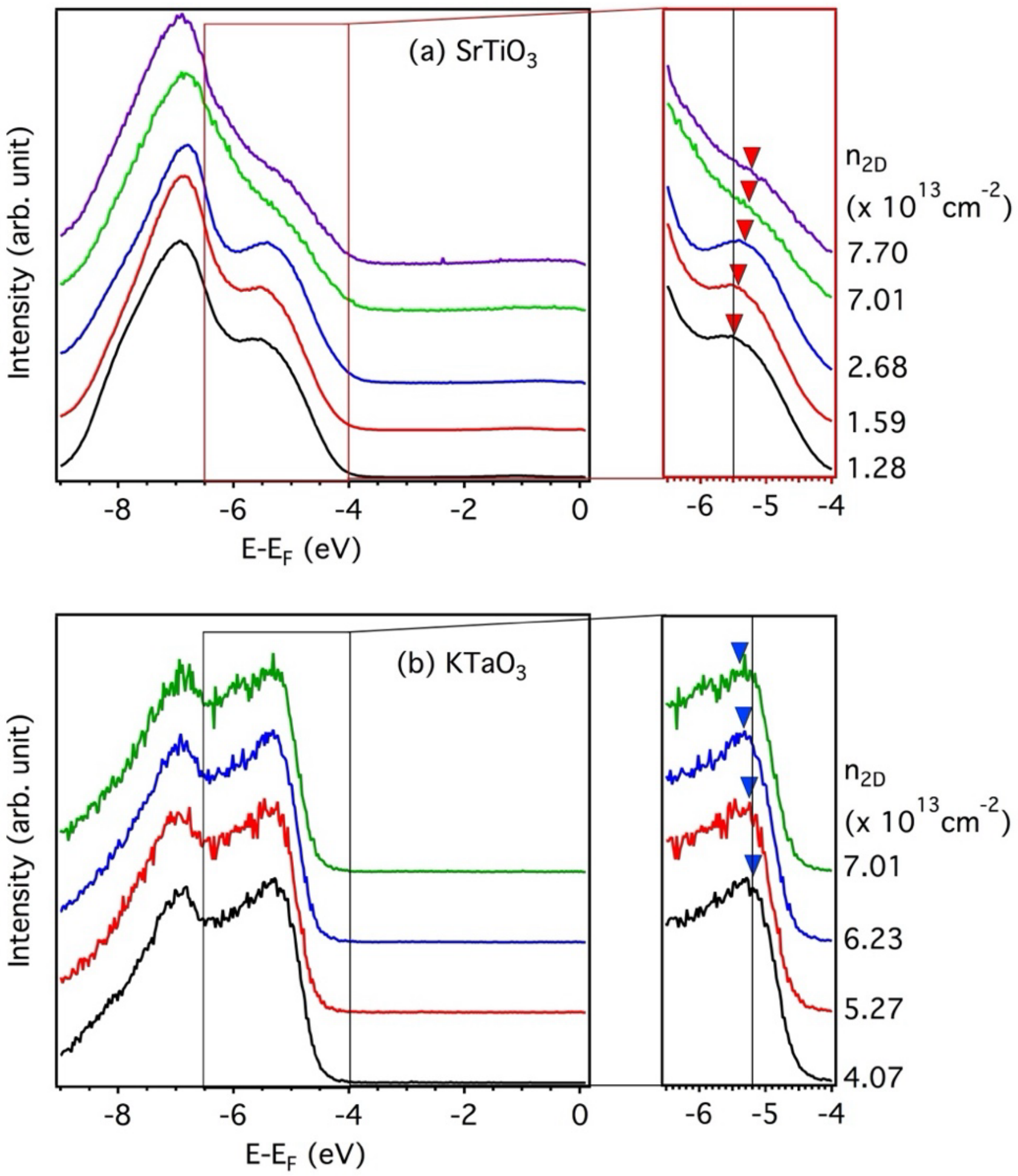

**Fig. S6** Comparison of the valence band evolution of (a) $SrTiO_3$ and (b) $KTaO_3$ as a function of electron density. Red and blue triangle symbols indicate spectral peaks of valence band for $SrTiO_3$ and $KTaO_3$, respectively.